\documentclass{iopart}

\usepackage{hyperref}
\usepackage{color,graphicx}
\usepackage{iopams}

\begin{document}

\title[The dark matter halo velocity anisotropy]{What it takes to measure a fundamental difference between dark matter and baryons: the halo velocity anisotropy}
\date{\today}
\author{Ole Host, Steen H Hansen}
\address{Dark Cosmology Centre, Niels Bohr Institute, University of Copenhagen, Juliane Maries Vej 30, DK-2100 Copenhagen \O{}, Denmark}
\ead{olehost@dark-cosmology.dk}

\begin{abstract} Numerous ongoing experiments aim at detecting WIMP dark matter particles from the galactic halo directly through WIMP-nucleon interactions. Once such a detection is established a confirmation of the galactic origin of the signal is needed. This requires a direction-sensitive detector. We show that such a detector can measure the velocity anisotropy $\beta$ of the galactic halo. Cosmological N-body simulations predict the dark matter anisotropy to be nonzero, $\beta\sim0.2$. Baryonic matter has $\beta=0$ and therefore a detection of a nonzero $\beta$ would be strong proof of the fundamental difference between dark and baryonic matter. We estimate the sensitivity for various detector configurations using Monte Carlo methods and we show that the strongest signal is found in the relatively few high recoil energy events. Measuring $\beta$ to the precision of $\sim0.03$ will require detecting more than $10^4$ WIMP events with nuclear recoil energies greater than $100\,$keV for a WIMP mass of $100\,$GeV and a $^{32}$S target. This number corresponds to $\sim10^6$ events at all energies. We discuss variations with respect to input parameters and we show that our method is robust to the presence of backgrounds and discuss the possible improved sensitivity for an energy-sensitive detector. 
\end{abstract}
\maketitle

\section{Introduction}
Several cosmological probes have shown that there is a dark matter component in the universe comprising about one quarter of the average energy density \cite{Spergel:2006hy,Cole:2005sx,Tegmark:2003uf}. The local density of dark matter in the solar neighborhood is estimated to be in the range $\rho_0=0.2-0.6\,$GeV/cm$^3$ \cite{Jungman:1995df}. Several dark matter candidates in the form of WIMPs, Weakly Interacting Massive Particles, have been suggested including the lightest supersymmetric particle (LSP) in SUSY extensions of the standard model, sterile neutrinos and axions -- see the review \cite{Bertone:2004pz} for a discussion. 

Recent years have seen a remarkable progress in the understanding of dark matter structures. Universal trends have been identified and quantified using numerical cosmological simulations.
One of the most discussed general trends lies in the behaviour of the universal density profile
\cite{Navarro:1996gj,Moore:1997sg,Fukushige:1996nr},
which has likely been explained \cite{Hansen:2004gs,Austin:2005ks,Dehnen:2005cu}.
Another general result of cosmological simulations is that the velocity anisotropy is
zero near the central region, and positive in the outer region \cite{Carlberg:1997qe,Cole:1995ep}.
There even appears to be a universal connection between the local slope of the
density profile and the local velocity anisotropy, which allows one to predict that
the value of the velocity anisotropy near the Earth should be non-zero and of the order 0.2
\cite{Hansen:2004qs,Hansen:2005cn}. Studies have shown that the velocity anisotropy can have a measurable effect on the detection rates of WIMPs \cite{Vergados:1999sf,Evans:2000gr,Green:2002ht}.
The physical interpretation of this velocity anisotropy is that the local dark matter `temperature'
is different in the tangential and radial directions with respect to the galactic centre. Thus, a non-zero velocity anisotropy
of the dark matter presents a sharp contrast with a typical baryonic gas.
This implies that an eventual
measurement of the anisotropy would be a strong proof that dark matter
really behaves significantly and fundamentally different from ordinary matter.
This is the main reason why we here present a possible method, based on direct detection of dark matter, by which one
eventually will be able to measure this property characterizing the dark matter halo. 

Detecting dark matter directly involves measuring the recoil of a nucleus which is scattered by the WIMP. At present, several direct detection experiments search for WIMPs from the galactic halo. Strategies differ, but the best exclusion limits at the moment are provided by low background cryogenic detectors detecting phonons and either ionization, such as CDMS \cite{Akerib:2005kh} and EDELWEISS \cite{Sanglard:2006hd}, or scintillation as CRESST \cite{Bravin:1999fc}. However, KIMS \cite{Lee:2007iq} are also competitive using only scintillation. A promising alternative is dual-phase noble gas detectors measuring scintillation and ionization, for example ZEPLIN \cite{Alner:2007ja}, XENON10 \cite{Baudis:2007ew} and ArDM \cite{Laffranchi:2007da}. These are relatively easily scalable to ton-mass detectors. The only collaboration to claim a detection so far is DAMA \cite{Bernabei:2000qi} which relies on detecting the weak annual modulation of the signal rate induced by the motion of the Earth \cite{Freese:1987wu}. However, the claim appears to be ruled out by CDMS \cite{Akerib:2005kh} and is heavily disputed. Of particular interest to the present work are the direction sensitive detectors DRIFT \cite{Alner:2005xp} and NEWAGE \cite{Miuchi:2007jy}. DRIFT is a $1\,$m$^3$ negative ion time projection chamber (TPC) situated in the Boulby Mine in the UK. The collaboration has provided proof-of-principle and are running the second stage of the detector. NEWAGE is a micro-TPC in the R\&D stage. 

\begin{figure}
\includegraphics[width=.5\textwidth]{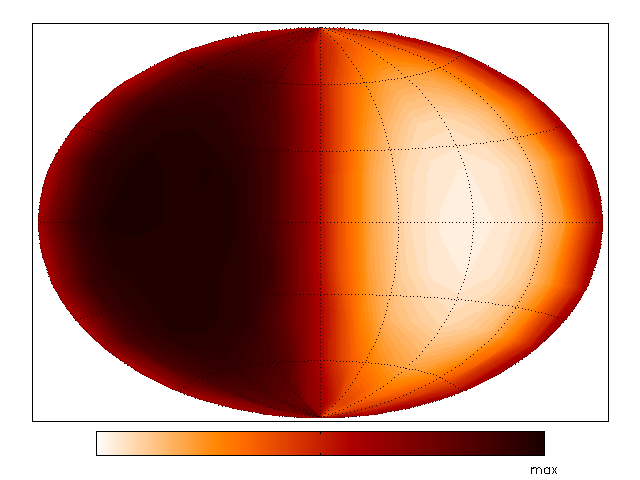}
\includegraphics[width=.5\textwidth]{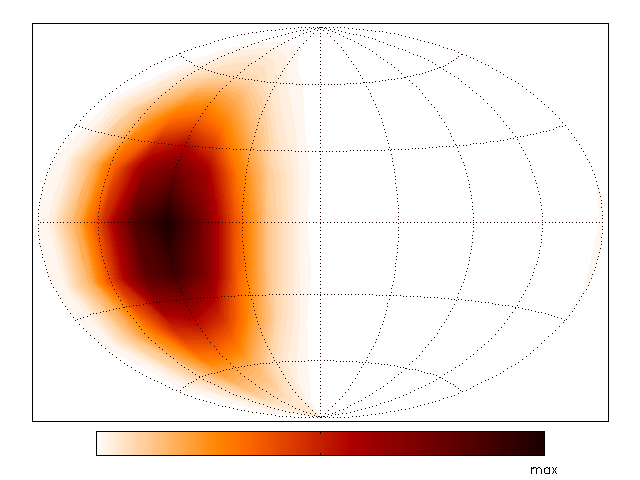}
\caption{Hammer-Aitoff projections of the sky in galactic coordinates showing the directional nuclear recoil rate in our fiducial model. On the left all recoils are shown, on the right only those with a recoil energy $E_{\mathrm{rec}}>100\,$keV. The Sun moves towards $90^{\circ}$ longitude on the Galactic equator.}\label{fi:ai}
\end{figure}

Direction-sensitive detectors search for a WIMP signal induced by the solar motion through the WIMP halo \cite{Spergel:1987kx}. This causes a large forward-backward asymmetry in the recoil signal rate as shown in figure \ref{fi:ai}. The magnitude of the asymmetry depends mainly on the solar orbital speed. Analyses suggest that less than 10 WIMP events in a direction-sensitive detector may confirm the signal as being galactic due to the large forward-backward asymmetry \cite{Copi:1999pw,Copi:2000tv,Green:2006cb,Morgan:2004ys}. Hence, while direction sensitive detectors are less likely to provide first detection, they may well  provide a crucial confirmation of the galactic origin of a WIMP signal.  However, the asymmetry is also weakly dependent on the velocity anisotropy of the dark matter halo and therefore a careful measurement of the asymmetry allows for a determination of $\beta$.

\section{Modelling}
In this section we discuss the components of the Monte Carlo simulation. The simulation generates a detector event by randomly selecting a WIMP velocity in the galactic halo, transforming the velocity to the detector and calculating the detector response to the event. 

\subsection{Dark matter velocity distribution}
We make the simplest assumption for the velocity distribution which is a modified Maxwell-Boltzmann distribution,
\begin{equation}
f(\vec{v})\,\rmd^3v=\frac{1}{(2\pi)^{3/2}\sigma_r^{\phantom{t}}\sigma_t^2}\exp\left(-\frac{v_r^2}{2\sigma_r^2}-\frac{v_\theta^2+v_\phi^2}{2\sigma_t^2}\right)\mathrm{d}^3v,
\end{equation}
in which the one-dimensional radial and tangential velocity dispersions are related through $\beta=1-\sigma_t^2/\sigma_r^2$. As explained in the introduction, numerical simulations predict $\beta\sim0.2$ near the Earth. When varying $\beta$ we fix the total velocity dispersion  $\sigma=\sqrt{\sigma_r^2+2\sigma_t^2}$ at $\sigma=v_0=230\,\mathrm{kms}^{-1}$, the circular orbit velocity. The distribution is cut off at an escape velocity of $v_{\mathrm{esc}}=600\,\mathrm{kms}^{-1}$. More general distributions \cite{Hansen:2005yj,vergados} will require a more detailed study.

The velocity of the Earth through the Galaxy is calculated as specified in Appendix B of \cite{Lewin:1995rx} to obtain the WIMP velocity $\vec{v}$ in the detector. The velocity is the sum of the Earth's orbit around the Sun $\vec{v}_E$, the solar motion $\vec{v}_\odot$ with respect to the local standard of rest and the galactic orbital speed of the local standard of rest, which is just the circular orbit velocity $v_0$. The solar motion measured by the Hipparcos satellite \cite{Dehnen:1997cq} is, in Galactic coordinates, $\vec{v}_\odot=(10.0,5.2,7.2)\,$kms$^{-1}$. We refer to \cite{Lewin:1995rx} for the calculation of $\vec{v}_E$ as a function of time and date.

\subsection{WIMP-nucleus interaction}
A WIMP can be detected by the nuclear recoil produced when it scatters off a target nucleus in a detector. We consider the case where the dominant channel is spin-independent elastic scattering, which is coherently enhanced by the number of nucleons at low energies. The cross section depends on the momentum transfer $q$ as
\begin{equation}
\sigma(q)=\sigma_0 |F(q)|^2,
\end{equation}
where $F(q)$ is the nuclear form factor and $\sigma_0$ is the cross section in the limit of zero momentum transfer. Following \cite{Lewin:1995rx}, the WIMP-nucleus interaction is modelled by the Helm form factor \cite{PhysRev.104.1466},
\begin{equation}\label{eq:ff}
F(qr_n)=3\frac{j_1(qr_n)}{qr_n}\exp{(-(qs)^2/2)},
\end{equation}
where $r_n=1.14A^{1/3}$ is the approximate nuclear radius and $s=0.9\,$fm is the skin thickness parameter. For elastic scattering the momentum transfer is given by
\begin{equation}\label{eq:q}
q=2\mu v\cos\theta, 
\end{equation}
where $\mu=m_tm_W/(m_t+m_W)$ is the reduced mass of the WIMP-target system, $v$ is the laboratory speed of the incoming WIMP and $\theta$ is the recoil angle of the nucleus with respect to $\vec{v}$. For each event, the maximum possible momentum transfer is determined by the velocity of the WIMP. The distribution $|F(q)|^2$ is then randomly sampled in the interval up to the maximum momentum transfer to determine the actual momentum transfer. This fixes the scattering angle $\theta$ through (\ref{eq:q}).

\subsection{Detector} 
The directional WIMP signal is unavoidably smeared out by the fact that only the nuclear recoil is observed, not the WIMP itself. In addition to this, the detector will have a limited angular resolution with which the initial nuclear recoil direction can be reconstructed. For example, in the DRIFT time projection chamber, the ionization cloud from a typical recoil will drift onto only a few anode wires in the readout \cite{spooner}. This, coupled with the charge diffusion, necessarily limits the accuracy with which the recoil can be reconstructed. High energy recoil events produce longer ionization tracks, hence it is easier to measure the direction of more energetic events.

In an actual experiment the angular resolution must be carefully measured. Here, we model it by rotating the recoil velocity in a random direction by an angle $\alpha$ drawn from the Fisher distribution on the sphere $p(\alpha)\propto \exp(\kappa \cos\alpha)$  \cite{19530507}. The parameter $\kappa$ fixes the width of the distribution with larger $\kappa$ corresponding to a more centralized distribution. We consider perfect reconstruction of the direction, $\kappa=\infty$, as well as $\kappa=5$ and $\kappa=2.3$ corresponding to half the sampled angles being greater than $30^\circ$ and $45^\circ$, respectively. 

Another parameter characterizing the detector is the detector threshold energy $E_{\mathrm{thr}}$. A realistic detector cannot measure the direction of nuclear recoils below this energy since low energy recoils will not move sufficiently long distances in the detector. Naturally the total signal rate is lowered for increasing threshold energy. 

\subsection{Calibrating the asymmetry-anisotropy relation}
The observed asymmetry depends on $\beta$ in a non-trivial way which necessitates Monte Carlo calibration. 
\begin{figure}
\includegraphics[width=\textwidth]{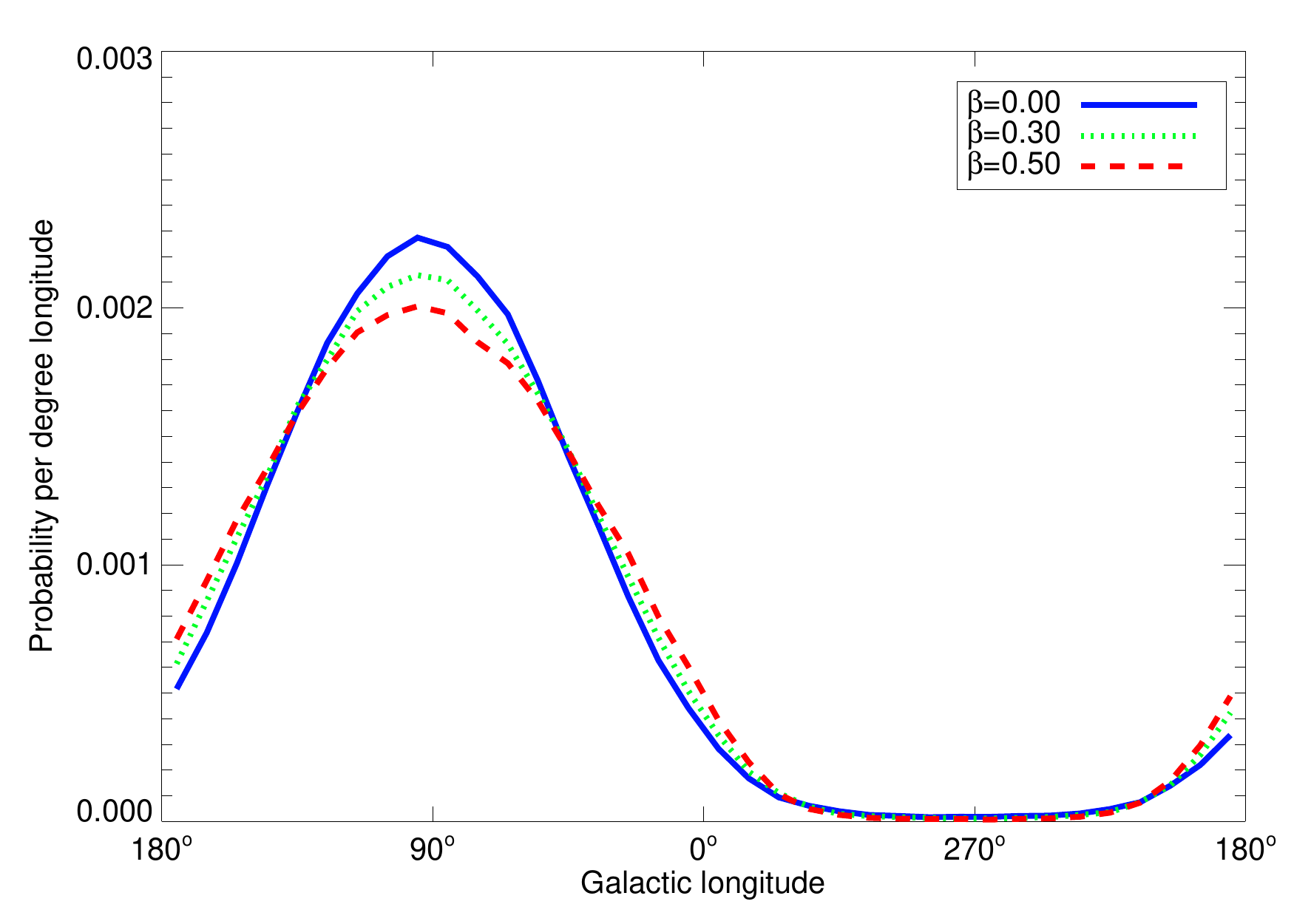}
\caption{Distributions of nuclear recoils with respect to galactic longitude, normalized to unity, for three values of $\beta$. An energy threshold of $E_{\mathrm{thr}}=100\,$keV has been applied, other parameters are as in the fiducial model.}
\label{fi:len}
\end{figure}

We define the angle $\phi$ between the observed nuclear recoil direction and the direction of the solar motion through the halo. If $\phi$ is less than a chosen acceptance angle $\alpha_{\mathrm{acc}}$ the event is counted as a forward event while if $\phi$ is greater than $180^\circ-\alpha_{\mathrm{acc}}$ the event is a backward event. The observed asymmetry is the difference between the number of forward and backward events. From figure \ref{fi:len}, it is evident that the asymmetry is large, compared to the total number of signals, but the dependence on $\beta$ is weak. The relative asymmetry is the difference between the number of forward and backward events with $E_{\mathrm{rec}}>E_{\mathrm{thr}}$, divided by the total number of generated events, regardless of recoil energy. In other words, it is the probability that a random detector event will add to the asymmetry given $E_{\mathrm{thr}}$ and $\alpha_{\mathrm{acc}}$.

\begin{figure}
\includegraphics[width=.5\textwidth]{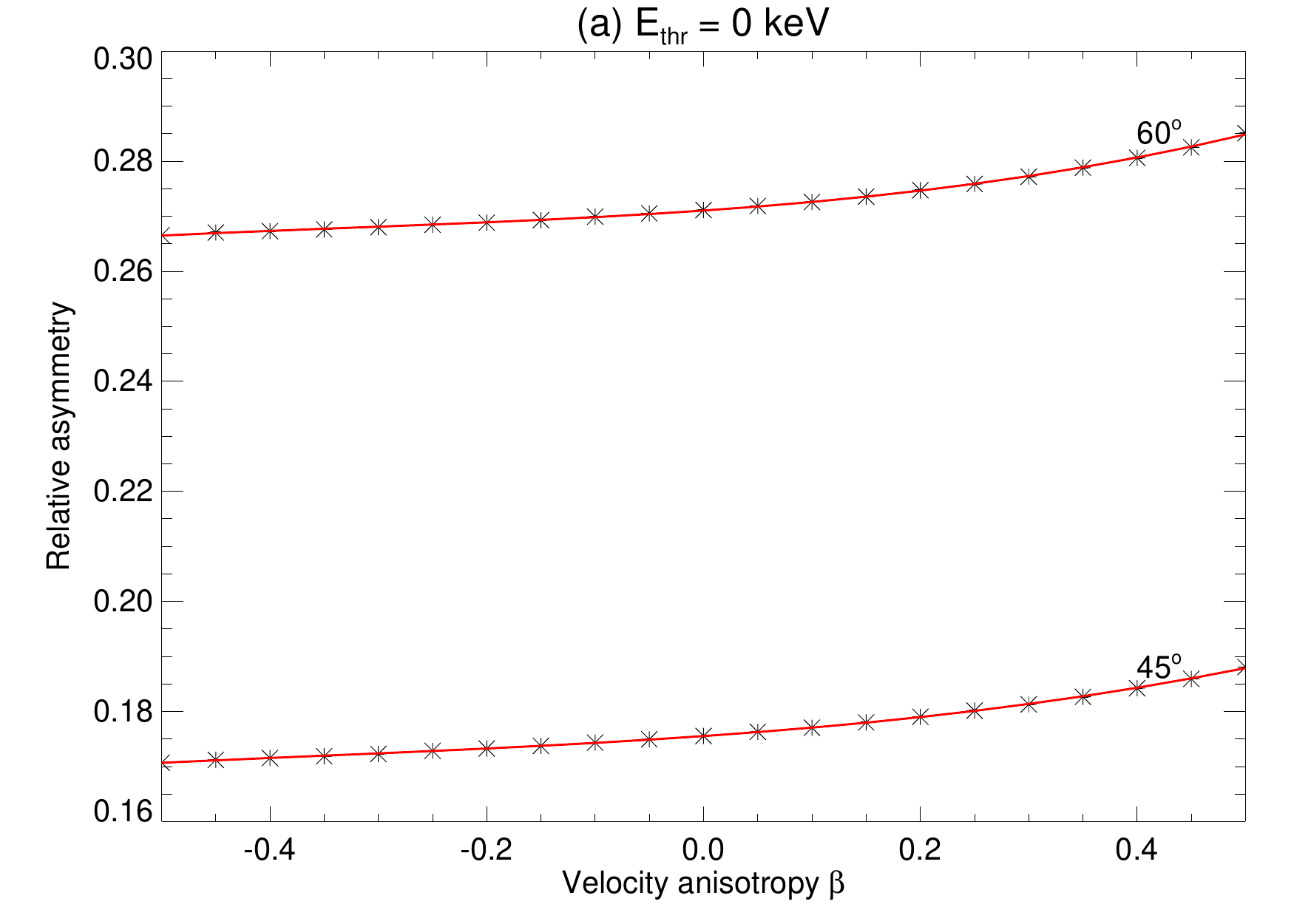}
\includegraphics[width=.5\textwidth]{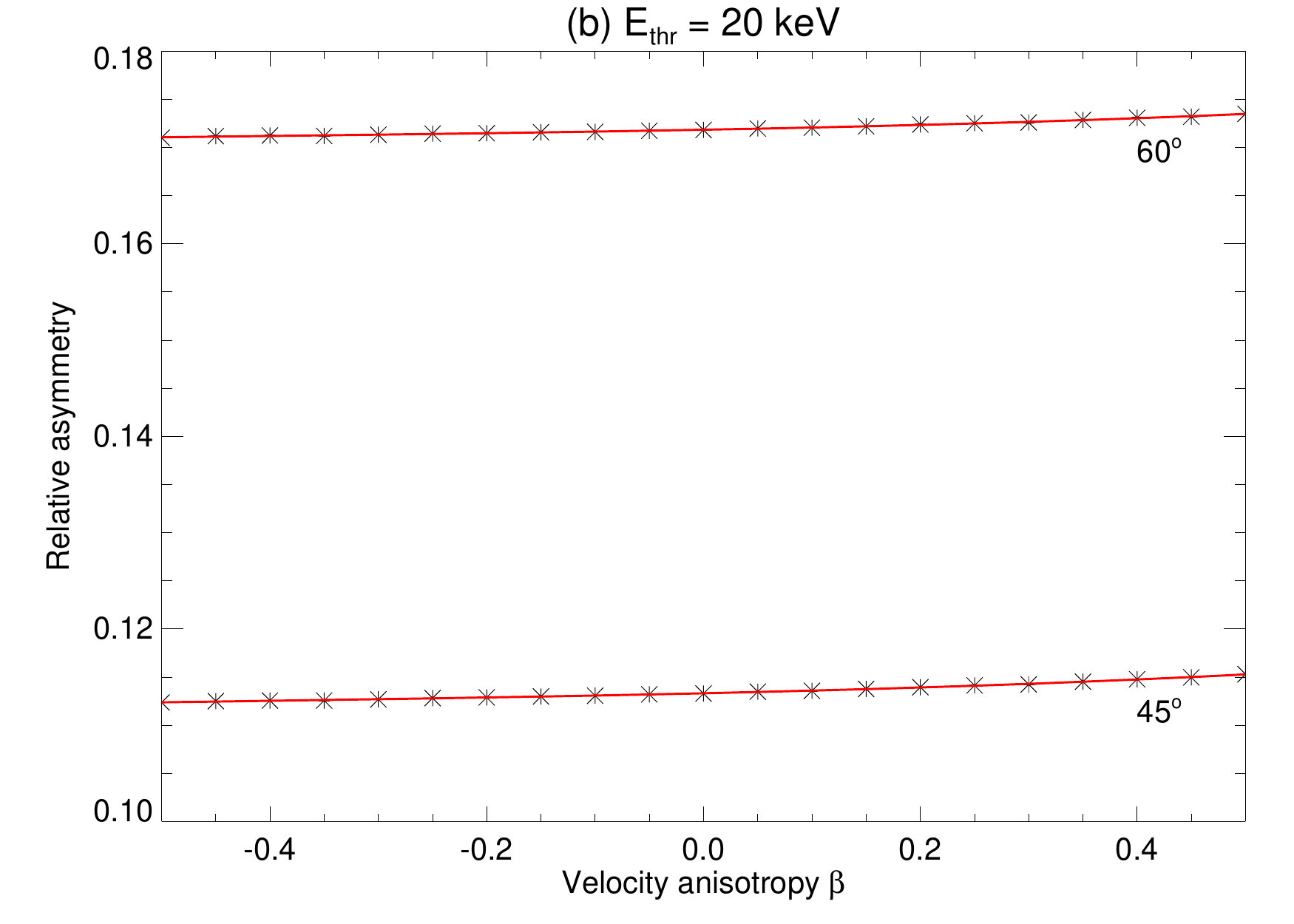}
\linebreak
\includegraphics[width=.5\textwidth]{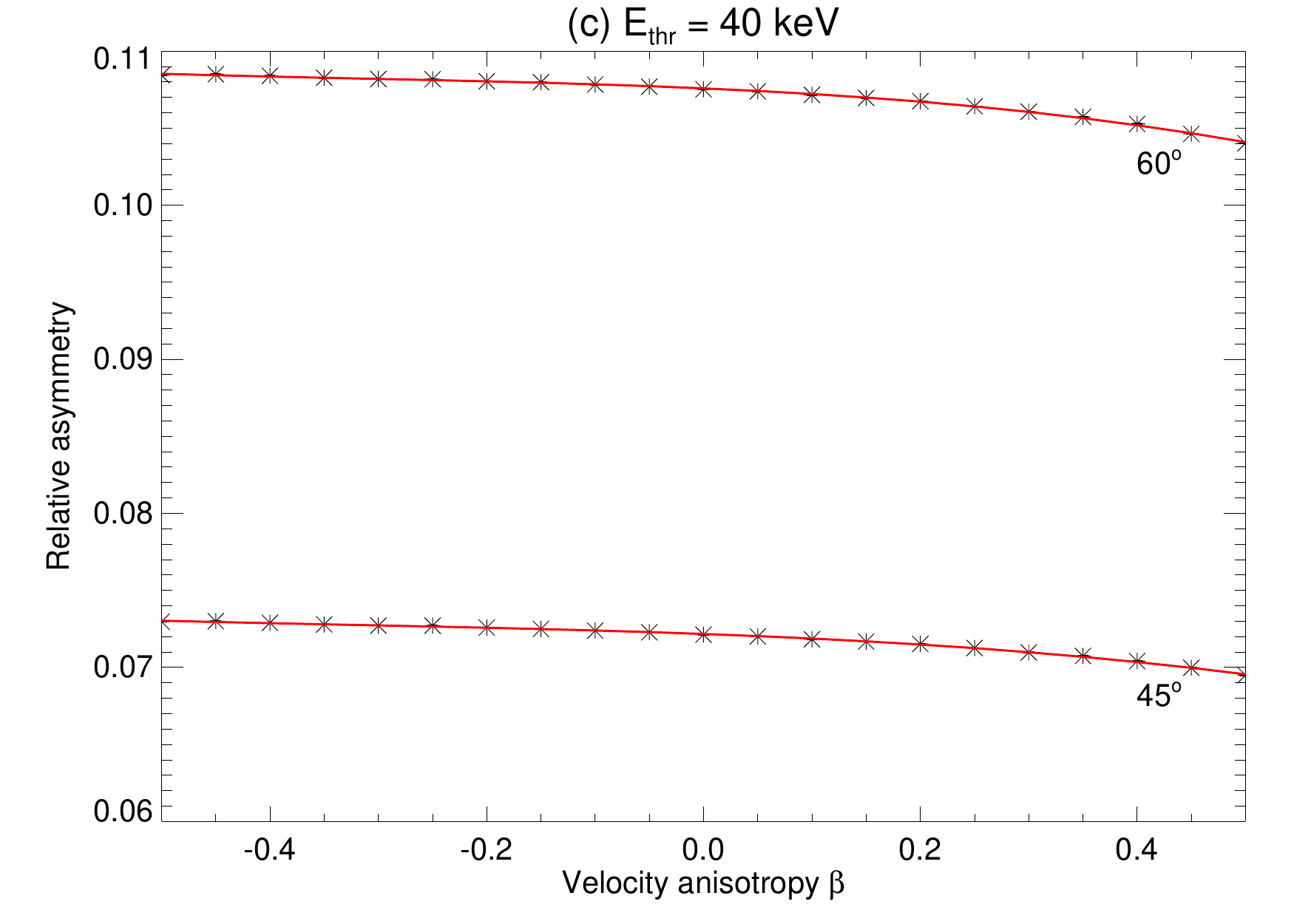}
\includegraphics[width=.5\textwidth]{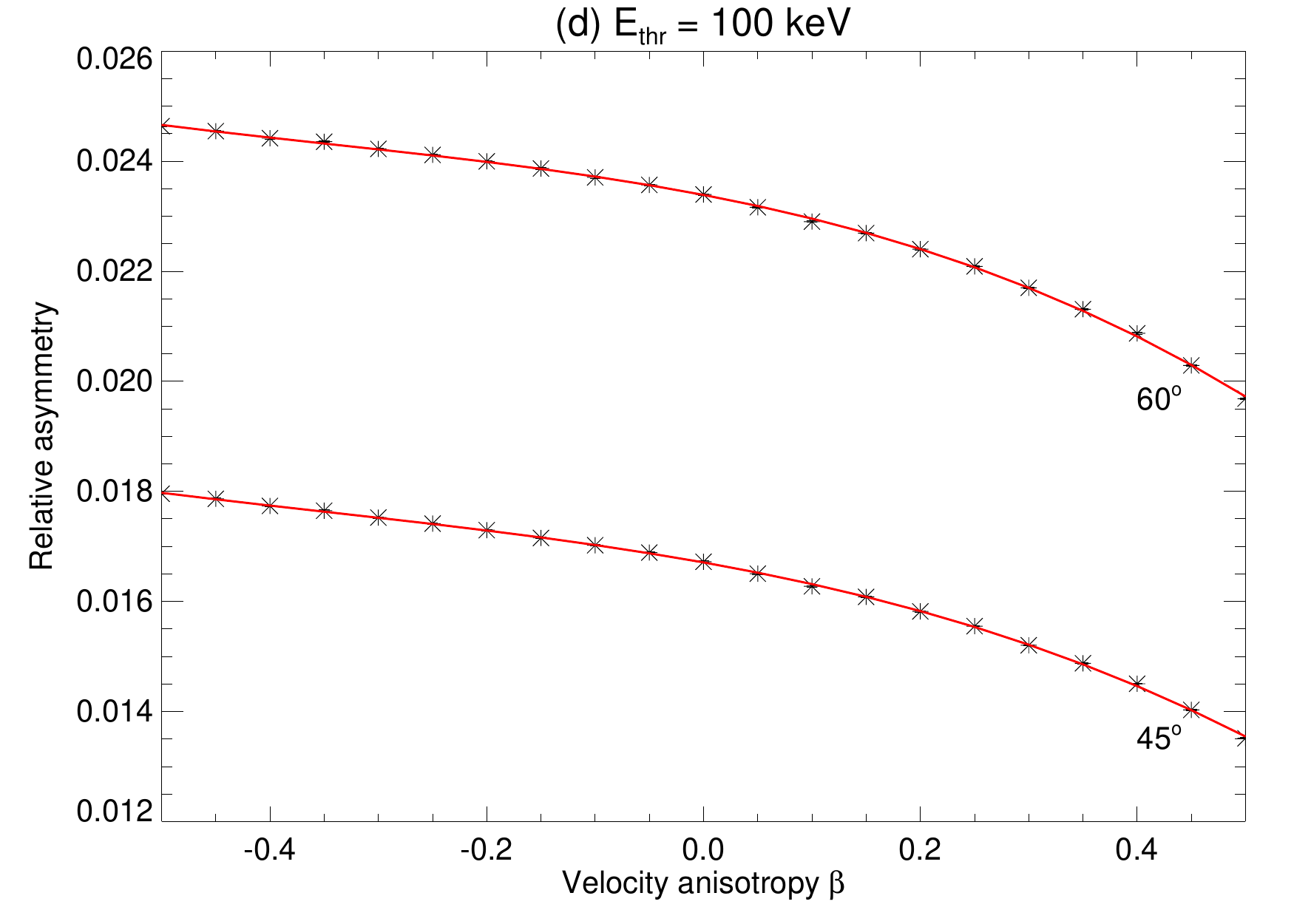}
\caption{Calibration curves for (a) $E_{\mathrm{thr}}=0\,$keV, (b) $E_{\mathrm{thr}}=20\,$keV, (b) $E_{\mathrm{thr}}=40\,$keV and (d) $E_{\mathrm{thr}}=100\,$keV, acceptance angles as labelled. The calibration is for the fiducial model, i.e.~$\kappa=\infty$, $m_W=100\,$GeV and $^{32} $S target.}\label{fi:cal}
\end{figure}

The relation between the relative asymmetry and the velocity anisotropy is calibrated by simulating $10^8$ events for $\beta$ in the range $(-0.5,0.5)$ in steps of $0.05$. For each $\beta$, the relative asymmetry is tabulated for a number of threshold energies and acceptance angles.  Third order polynomials are fitted to the calculated relative asymmetries and used as calibration curves. Figure \ref{fi:cal} shows the calibration curves at threshold energies of $0\,$keV and $100\,$keV. Note that for low threshold energy, the relative asymmetry increases with $\beta$ while the opposite is true for high threshold energy. This change in behaviour occurs in the region of $E_{\mathrm{thr}}\simeq40\,$keV, increasing with the smaller acceptance angles.

The calibration depends on both the WIMP and target masses, $m_W$ and $m_t$, on the assumed velocity dispersion $\sigma$, and on the angular resolution of the detector, $\kappa$. Further, each configuration of detection energy threshold and acceptance angle results in a different calibration. We calculate calibration curves for $E_{\mathrm{thr}}/(\mathrm{keV})\,\in\left[0,20,40,70,100,140,180\right]$ and $\alpha_{\textrm{acc}}\in\left[90^\circ,60^\circ,45^\circ,36^\circ,30^\circ,26^\circ,22^\circ\right]$.

\section{Results}\label{sc:res}
Now we discuss the sensitivity, i.e.~the mean or expected accuracy, with which the velocity anisotropy can be measured, depending on the experimental configuration. We simulate a large number of experiments, each measuring $\beta$ from a number of observed nuclear recoils. The measurement is done by converting the observed relative asymmetry in each experiment to a value for $\beta$ using the calibration curves discussed above. This yields a distribution of measurements and the sensitivity is the width of this distribution.

\subsection{Signal rates}
We do not calculate specific rates, rather we assume a WIMP mass and a target nucleus and estimate the sensitivity from a number of generated events for different $E_{\mathrm{thr}}$ and $\alpha_{\mathrm{acc}}$. Increasing $E_{\mathrm{thr}}$ lowers the signal rate by a factor depending on the masses. Table \ref{tb:rel} lists the fraction of recoils with energy above a given threshold for some combinations of $m_W$ and target nucleus. For example, out of 1000 recoils only about 30 would have recoil energy greater than $100\,$keV if $m_W=100\,$GeV and the target is $^{32}$S. It is therefore natural to compare sensitivities for different threshold energies between simulations with the same number of total generated events.
\fulltable{\label{tb:rel}Fraction of nuclear recoils with recoil energy above $E_{\mathrm{thr}}$.}\br
$E_{\mathrm{thr}}$ (keV) & 0 & 20 & 40 & 70 & 100 & 140 & 180 \\ \mr
$^{32}S$ target, $m_W=100\,$GeV & 1.0 & 0.35 & 0.18 & 0.074 & 0.030 & 0.0082 & 0.0018 \\
$^{12}C$ target, $m_W=100\,$GeV & 1.0 & 0.22 & 0.073 & 0.013 & 0.0014 & - & - \\
$^{32}S$ target, $m_W=500\,$GeV & 1.0 & 0.44 & 0.27 & 0.14 & 0.073 & 0.031 & 0.013 \\
\br \endfulltable

The actual signal rate in a detector is proportional to $\rho_0/m_W$, where $\rho_0$ is the WIMP density in the vicinity of the solar system.

\subsection{Fiducial model}
\begin{figure}
\includegraphics[width=\columnwidth]{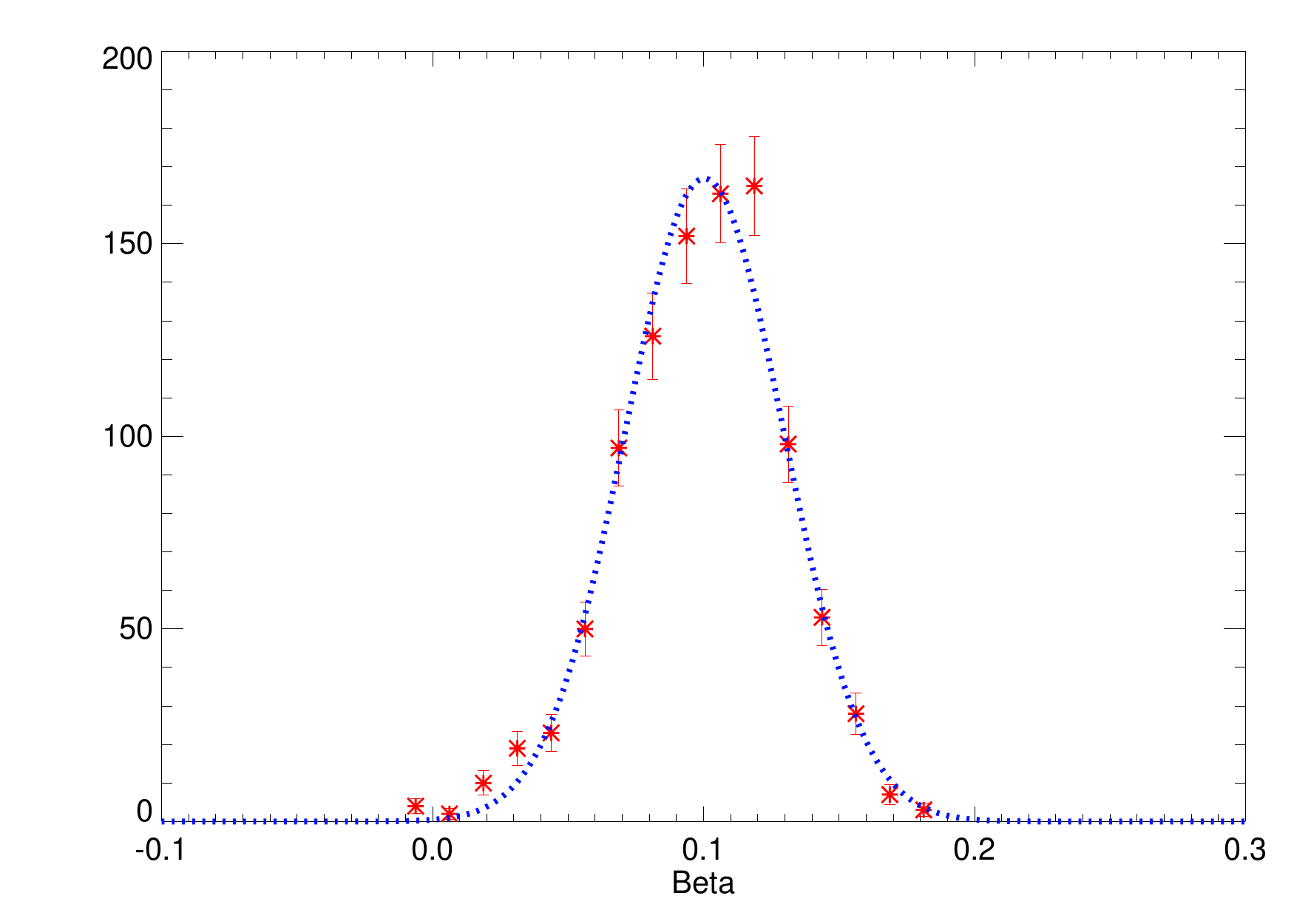}
\caption{Binned distribution of measured $\beta$'s in the fiducial model with $10^6$ detector events, $E_{\mathrm{thr}}=100\,$keV and $\alpha_{\textrm{acc}}=45^\circ$. The width of the distribution defines the sensitivity for this experimental configuration. The curve shows the Gaussian fit which has $\sigma_\beta=0.029$.}
\label{fi:his}
\end{figure}

We consider a fiducial model with velocity anisotropy $\beta=0.1$, a WIMP mass of $m_W=100\,$GeV, a $^{32}$S target nucleus and perfect reconstruction of the recoil direction, $\kappa=\infty$. We simulate a total of $10^9$ detector events which we group as 1000 experiments obtaining $10^6$ WIMP detections each. The distribution of measured $\beta$'s for these 1000 simulated experiments is shown in figure \ref{fi:his} for $E_{\mathrm{thr}}=100\,$keV and $\alpha_{\mathrm{acc}}=45^\circ$ . The width $\sigma_\beta$ of the distribution is the desired estimate of the experimental sensitivity to $\beta$. Explicitly, a detector reconstructing the direction of the roughly $3\times10^4$ events with recoil energy greater than $100\,$keV (out of the $10^6$ events at all energies) is expected to be able to measure $\beta$ with an accuracy of $\sigma_\beta=0.029$, if $\beta=0.1$.

\begin{figure}
\includegraphics[width=\columnwidth]{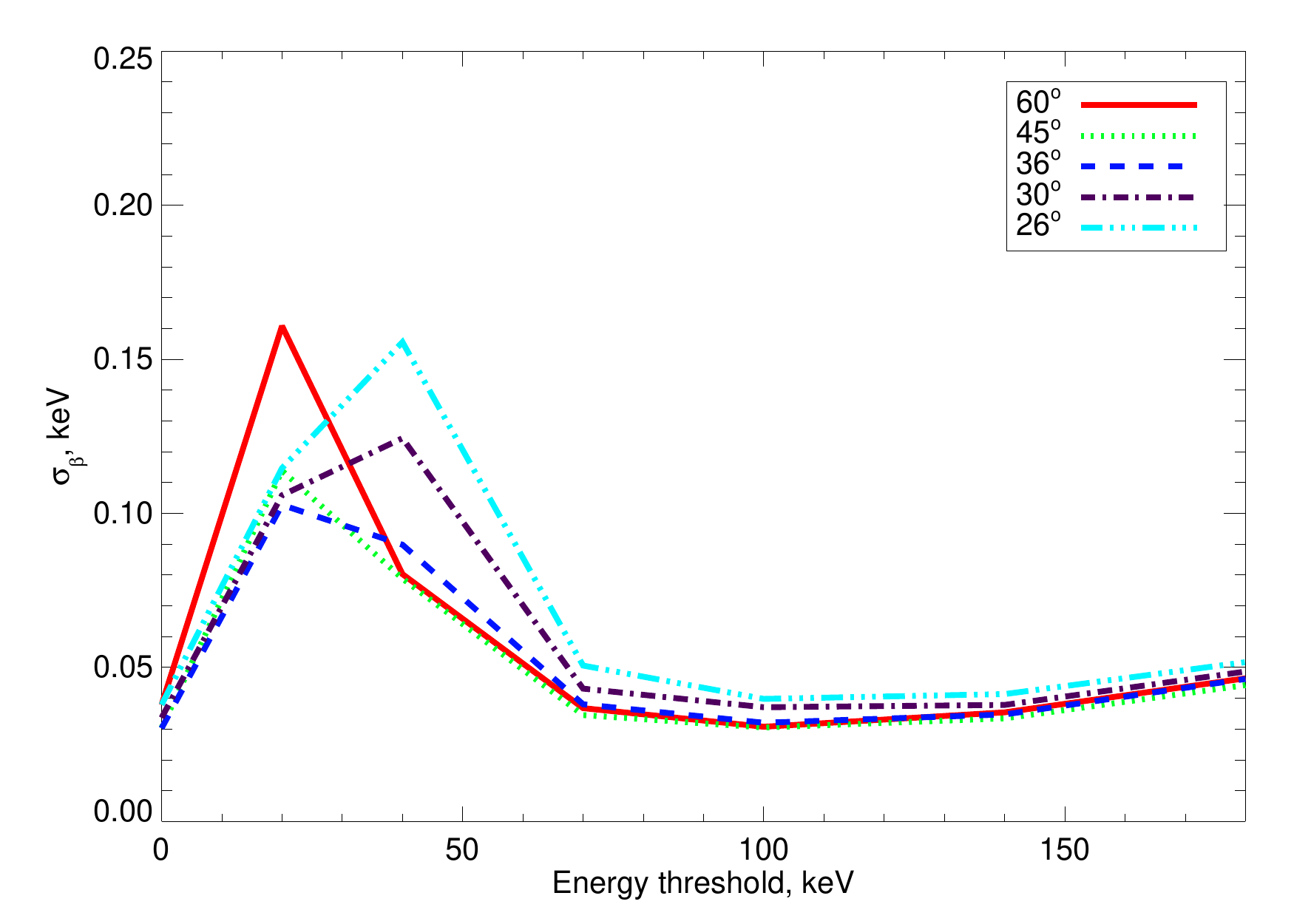}
\caption{Sensitivity $\sigma_\beta$ of measurement of $\beta$ as a function of detector energy threshold for various acceptance angles in the fiducial model.}\label{fi:sig}
\end{figure}

The sensitivity in the fiducial model is shown in figure \ref{fi:sig} for various acceptance angles as a function of threshold energy. The best sensitivity is obtained for the above-mentioned configuration, $\alpha_{\mathrm{acc}}=45^\circ$ and $E_{\mathrm{thr}}=100\,$keV, for which $\sigma_\beta=0.029$. 
Very similar results are obtained for $E_{\mathrm{thr}}=0\,$keV while intermediate threshold energies are less optimal. The dependence on $\alpha_{\mathrm{acc}}$ is weak for large threshold energies, but rather strong at low and intermediate $E_{\mathrm{thr}}$. The optimal threshold energy at $100\,$keV should be understood as the best compromise between a steep calibration curve and a large number of events. Explicitly, the asymmetry is more enhanced if only high energy events are detected, but the lower number of events reduces the signal to noise ratio. For zero threshold energy, the opposite happens as there is a large number of events but each event contains less information. The poor performance at intermediate $E_{\mathrm{thr}}$ is mainly due to the calibration curves changing from positive to negative slope as the threshold energy increases while the increase in $\sigma_\beta$ for $E_{\mathrm{thr}}>100\,$keV is due to the low number of events with sufficient recoil energy. The simulated experiments reproduce the input $\beta=0.1$ consistently. The only unfortunate exceptions are for $\alpha_{\mathrm{acc}}=90^\circ$ and  $E_{\mathrm{thr}}=20$ or $40\,$keV in which case the flatness of the calibration curves and the low signal per event smear the distribution out over a wide range of $\beta$'s.

The measured value of the relative asymmetry in the 1000 experiments is close to Gaussian since the number of forward events is very large, the number of backward events is very small and each number is Poisson distributed. However, since the calibration curves are not straight lines the distribution of measured $\beta$'s is distorted. The measure of this distortion is the sample skewness, which is the ratio of the third sample moment to the second,
\begin{equation}
\gamma=\frac{\mu_3}{\sigma^3}=
\frac{\sqrt{n}\sum_i(\beta_i-\bar{\beta})^3}{[\sum_i(\beta_i-\bar{\beta})^2]^{3/2}}.
\end{equation}
In particular, we find large, negative skewness $\gamma\lesssim-1$ for low threshold energies as well as for $\alpha_{\mathrm{acc}}=90^\circ$. In these cases a Gaussian fit is not a good representation of the distribution of $\beta$'s and we take the sample standard deviation of the unbinned distribution as the sensitivity instead.

\begin{figure}
\includegraphics[width=\columnwidth]{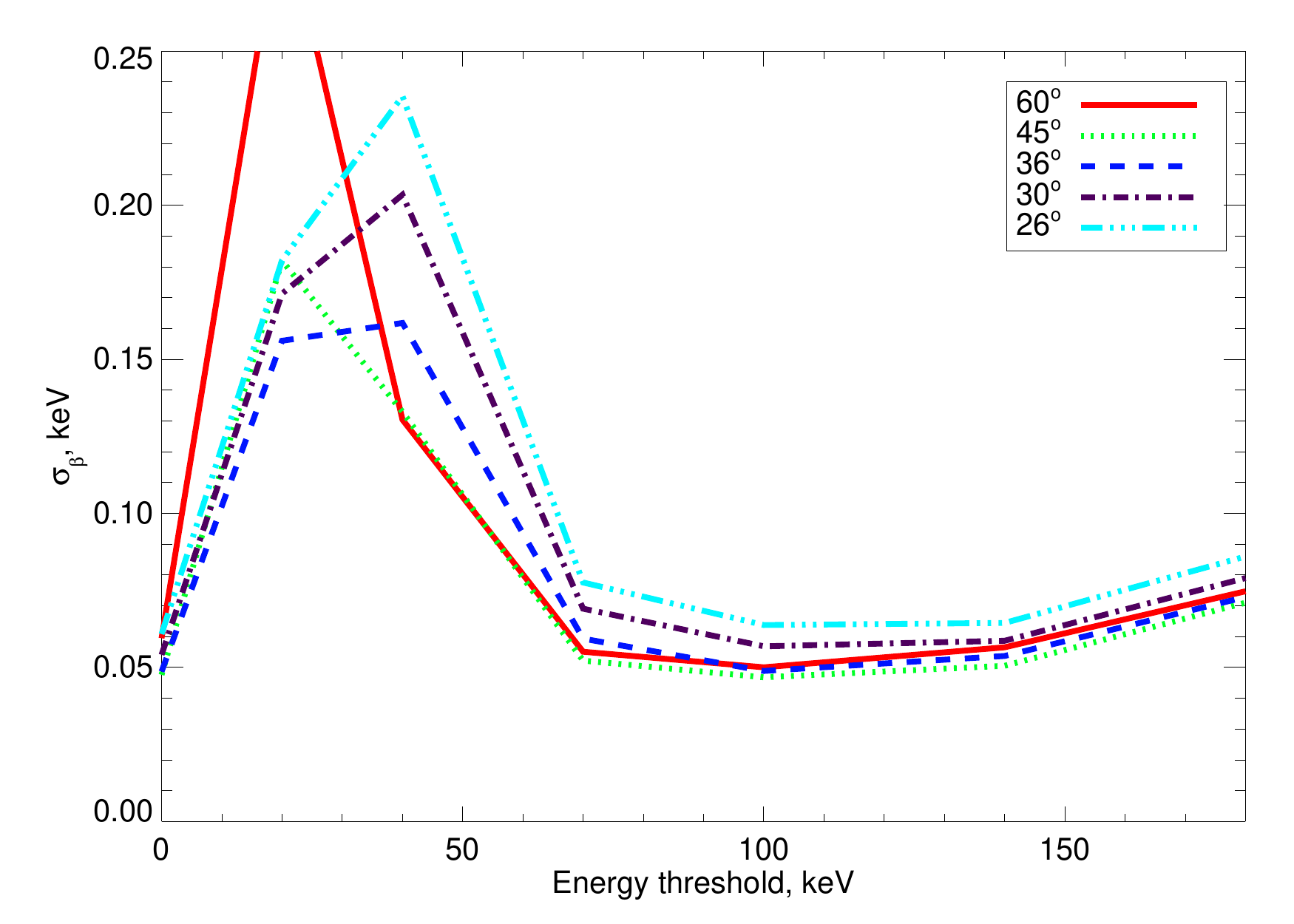}
\caption{Same as figure \ref{fi:sig} but with $4\times10^5$ events per experiment.}\label{fi:si2}
\end{figure}

The dependence on the number of detected events is investigated by regrouping the $10^9$ simulated events into more experiments with fewer detections. Figure \ref{fi:si2} shows the obtained accuracy for $4\times10^5$ events, i.e.~2.5 times fewer than the fiducial configuration. The best $\sigma_\beta$ is again obtained at $E_{\mathrm{thr}}=100\,$keV and $\alpha_{\mathrm{acc}}=45^\circ$, for which $\sigma_\beta=0.047$. In general we find that $\sigma_\beta\propto1/\sqrt{N}$ for experiments with at least $10^5$ events, as expected. However, the skewness becomes more and more pronounced as the number of detector events is reduced. 

In summary, to make a coarse measurement of the velocity anisotropy of the dark matter halo with an accuracy of about 0.1, about $10^5$ WIMPs at all energies would be needed. However, a detector which is insensitive to the recoil energy should ideally only count the nuclear recoils with energy greater than $100\,$keV. A precision measurement would require more than $10^6$ events, corresponding to a ton-scale direction-sensitive detector operating for several years if the cross-section is $10^{-7}$pb. These numbers, compared to the present status of WIMP directional detection experiments, are of course very large and would require a large dedicated experimental programme. Additionally, a cross-section of $10^{-7}\,$pb may be too optimistic -- for example, a recent study found that the most favored region would be the range $10^{-10}-10^{-8}\,$pb \cite{Trotta:2006ew}.

\subsection{Parameter influence}
Next, we investigate the consequences of varying each of the parameters of the fiducial model independently. 
\begin{itemize}
\item{First and most important, a limited detector angular resolution, $\kappa=5$, is found to increase $\sigma_\beta$ by 30\%. A poorer detector with $\kappa=2.3$ increases $\sigma_\beta$ by 75\%. The optimal acceptance angle in both cases is $\alpha_{\mathrm{acc}}=60^\circ$. In these cases, the $\sigma_\beta$'s obtained at zero threshold energy are notably poorer than at $100\,$keV, unlike in the fiducial model. One might expect the resolution would have a stronger effect but it should be remembered that the WIMP recoil distribution is already smeared out by the nuclear recoil distribution so the relative decrease due to a finite $\kappa$ is small.}
\item{If the target nucleus mass is lowered, the reduced mass of the WIMP-nucleus system is decreased and, from (\ref{eq:q}), so is the width of the recoil angle distribution. Hence the nuclear recoil direction resembles the incoming WIMP direction better. For a $^{12}$C target, $\sigma_\beta$ is decreased by 15\%. The lower reduced mass also lowers the optimal threshold energy to $E_{\mathrm{thr}}=70\,$keV.}
\item{If the WIMP mass is increased, the reduced mass is also increased and the recoil angle distribution becomes wider, resulting in poorer sensitivity. For $m_W=500\,$GeV we find $\sigma_\beta$ is increased by 10\% and the optimal threshold energy is $180\,$keV. It should be noted here that since the WIMP flux is inversely proportional to $m_W$, it would take five times as long to record the same number of events if $m_W=500\,$GeV rather than $100\,$GeV.}
\item{Finally, if the `true' value of $\beta$ is instead $-0.1$ the smaller slope of the calibration curve causes $\sigma_\beta$ to increase by about 50\%. The opposite is true if $\beta=0.3$, in which case $\sigma_\beta$ decreases by 40\%. Hence, if the actual value of $\beta$ is larger than the $0.1$ in our reference model, the number of events needed to measure $\beta$ is significantly smaller. For example, measuring $\beta=0.3$ to a precision of about $\pm0.06$ would require roughly $3\,000$ events above $100\,$keV. Figure \ref{fi:sb} shows the variation of the sensitivity with the assumed true value of $\beta$.
\begin{figure}[tbp]
\begin{center}
\includegraphics[width=\columnwidth]{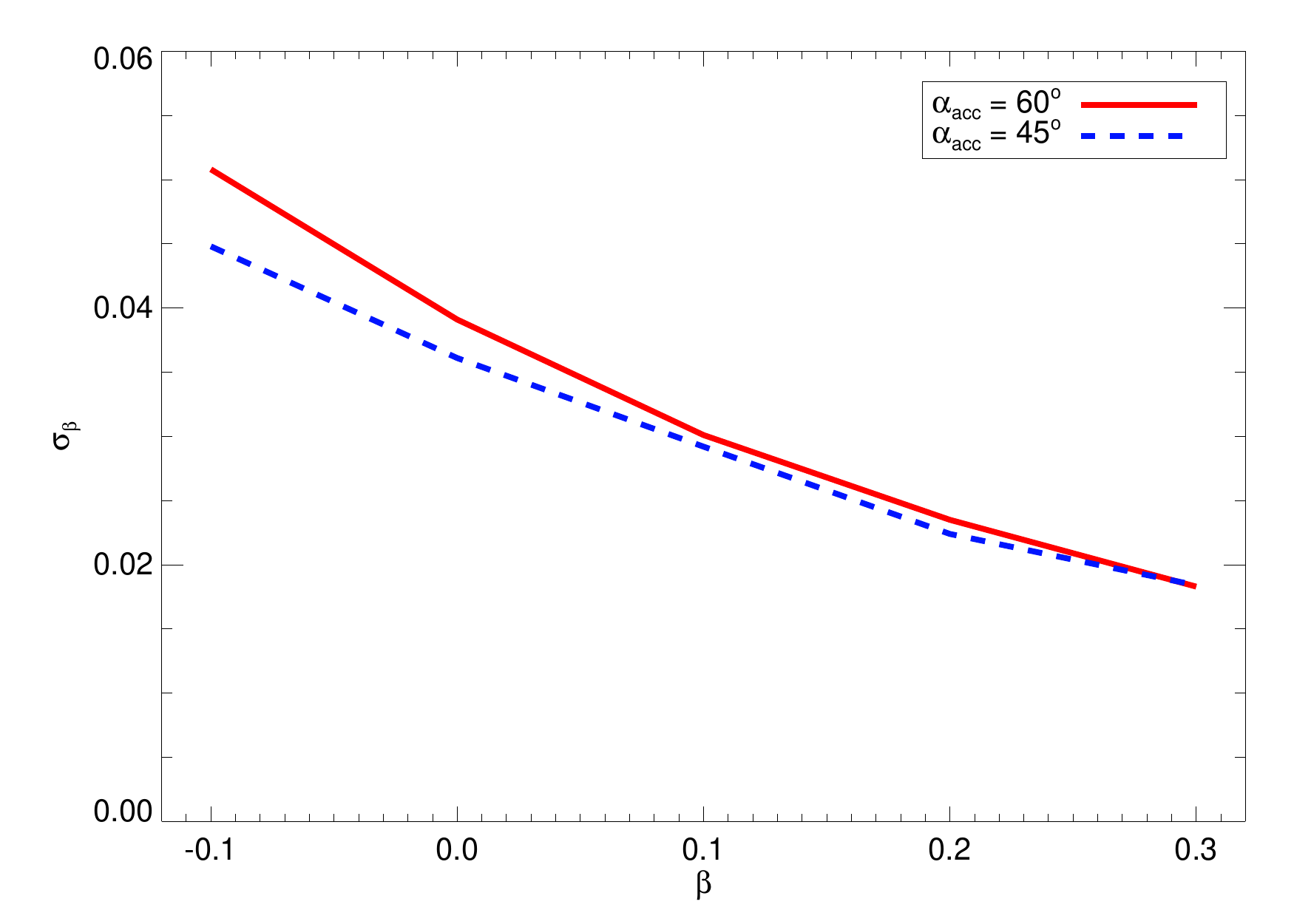}

\caption{Variation of the sensitivty with the input value of $\beta$ for $\alpha_\mathrm{acc}=45^\circ$ and $60^\circ$ and $E_{\mathrm{thr}}=100\,$keV. The dependence is primarily determined by the calibration curve.}
\label{fi:sb}
\end{center}
\end{figure}
} 
\end{itemize} 
The best obtained $\sigma_\beta$ for these parameters and the corresponding optimal acceptance angle and threshold energy are summarized in table \ref{tb:var}.
\Table{\label{tb:var}Impact of varying the parameters of the fiducial model independently.}\br
Parameter & Best $\sigma_\beta$ & $E_{\mathrm{thr}}$ (keV) & $\alpha_{\mathrm{acc}}$ \\ \mr
Fiducial model & $0.029$ & $100$ & $45^\circ$ \\
$\kappa=5$ & $0.037$ & $100$ & $60^\circ$ \\
$\kappa=2.3$ & $0.051$ & $100$ & $60^\circ$ \\
$m_W=500\,$GeV & $0.033$ & $180$ & $45^\circ$ \\
$^{12}$C target & $0.025$ & $70$ & $45^\circ$ \\
$\beta=-0.1$ & $0.045$ & $100$ & $45^\circ$ \\ \br
\endTable

\subsection{Background}
\begin{figure}
\includegraphics[width=\columnwidth]{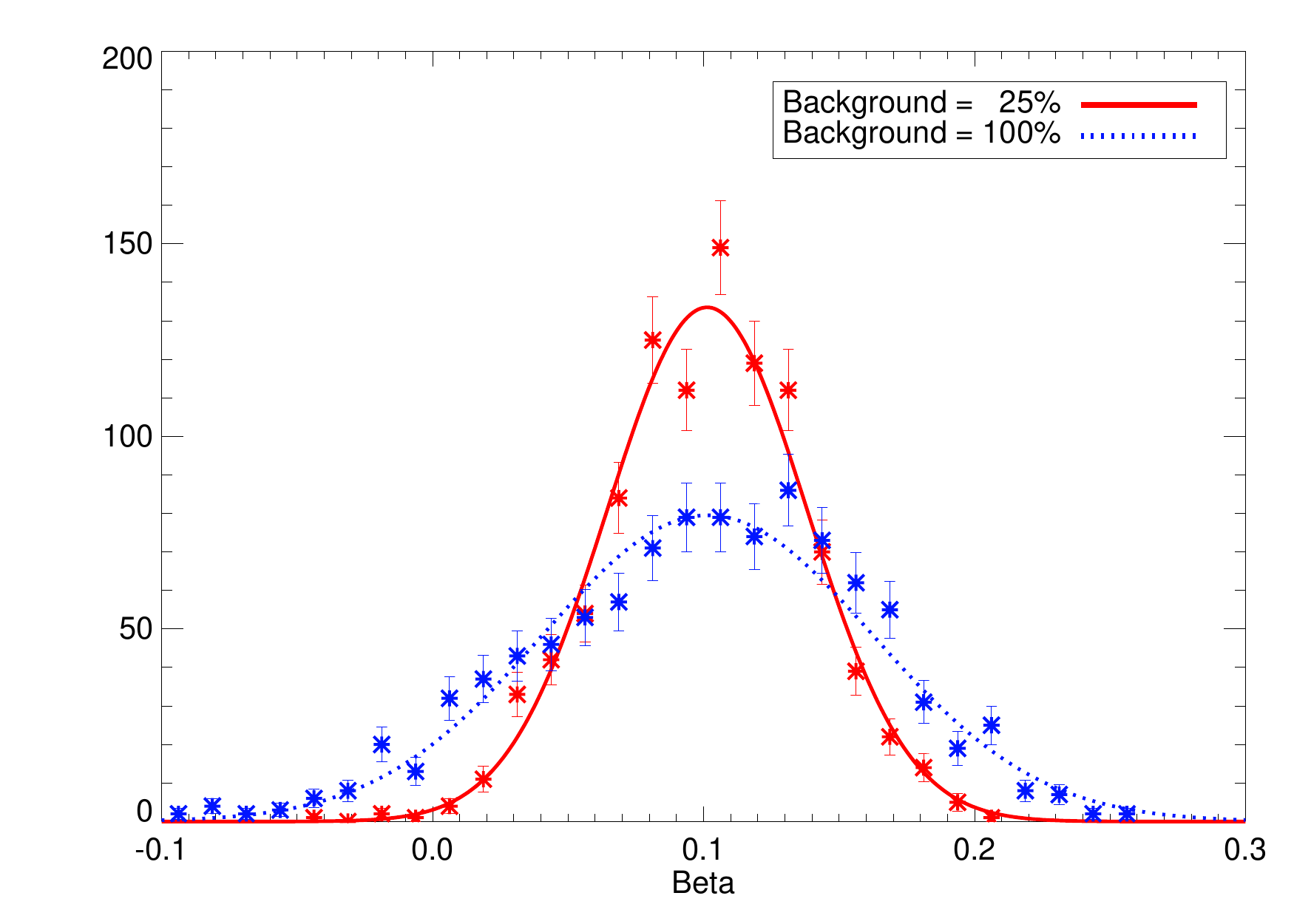}
\caption{Binned distributions of measured $\beta$'s in the fiducial model as in figure \ref{fi:his} but including a background level set relative to the number of forward events. The sensitivities are $\sigma_\beta=0.037$ and $\sigma_\beta=0.06$ for 25\% and 100\% background levels, respectively.}\label{fi:noi}
\end{figure}

So far we have assumed a zero background level. Now we discuss the influence of a nonzero background which is assumed to be isotropic in the Galactic frame when averaged over time \cite{Morgan:2005sq}. We consider the impact on the optimal detector configuration for the fiducial model, $E_{\mathrm{thr}}=100\,$keV and $\alpha_{\mathrm{acc}}=45^\circ$. For each simulated experiment we add a number of background events to the forward and backward signal events. These two numbers are drawn from a Poisson distribution with mean equal to a fraction of the number of forward signal events. The asymmetry is then calculated as before and the calibration curve is used to determine the measured value of $\beta$.  We find that the degradation due to nonzero background is benign, as long as the signal is not weaker than the background. For a 25\% background level the sensitivity $\sigma_\beta=0.037$, an increase of less than 30\%. For a 100\% background, $\sigma_\beta=0.06$ while for a 400\% background $\sigma_\beta=0.10$. The distribution of measured $\beta$'s still reproduce $\beta=0.1$ consistently as shown in figure \ref{fi:noi}. 

The well-behaved performance of the sensitivity with respect to nonzero backgrounds can be attributed to our method of measuring the velocity anisotropy from the difference of the number of forward and backward signals, defined within equal solid angles. This method is largely insensitive to the background level, since the expected number of background events is the same in the forward and backward bins. 

\subsection{Energy resolution}
We have not assumed that the detector can measure the nuclear recoil energy. However, as is evident from the calibration curves in figure \ref{fi:cal} the forward-backward asymmetry increases with $\beta$ for low energy recoils but decreases for high energy recoils. Hence a `cleaner' signal can be obtained if the experiment is able to place events into energy bins with some resolution. 

We have investigated the possible improvements by binning events into $15\,$keV bins according to their recoil energy, i.e.~with no additional detector resolution effects. For each energy bin and acceptance angle we calculate separate calibration curves. The flip from positive to negative slope of the calibration curves takes place at about $E_{\mathrm{rec}}=60\,$keV for the fiducial model. Hence, at intermediate recoil energies there is virtually no sensitivity to $\beta$. Following the usual procedure, we calculate the distribution of measured $\beta$'s in each energy bin. For $10^6$ events distributed over all bins, the measurements reproduce the input $\beta$ for low ($\lesssim50\,$keV) and high ($\gtrsim100\,$keV) recoil energies. The best sensitivity in individual bins is obtained in the lowest energy bin, $E_{\mathrm{rec}}\in(0,15)\,$keV for which $\sigma_{\beta,E}=0.032$. If the results of the bins that reproduce $\beta=0.1$ are combined through $\sigma_\beta^{-2}=\sum_i \sigma_{\beta,E_i}^{-2}$, a sensitivity of $\sigma_\beta=0.20$ is achieved. This corresponds roughly to the improvement obtained by our standard detector with no energy resolution taking data for twice as long.

It is interesting to note that a direction-sensitive detector with high energy resolution will be able to extend the study of \cite{Drees:2007hr} to extract the shape of the full 3-dimensional velocity distribution function.

\section{Discussion}
We have investigated the possibility of measuring the velocity anisotropy $\beta$ of the galactic dark matter halo in a direction sensitive WIMP detector using Monte Carlo methods. The measurement is based on the well known forward-backward asymmetry in the directional spectrum of WIMP-induced nuclear recoils. A non-zero $\beta$ alters the magnitude of the asymmetry slightly which makes it possible to measure $\beta$.

We find that in excess of $10^5$ events across all energies are needed to make a coarse measurement of $\beta$. An experiment measuring $3\times10^4$ events with recoil energy greater than $100\,$keV, equivalent to $10^6$ events at all energies, should be able to measure $\beta$ to a precision of $0.03$. This result is obtained for an acceptance angle of $\alpha_{\mathrm{acc}}=45^\circ$, a velocity dispersion of $\sigma=230\,$kms$^{-1}$, a WIMP mass of $100\,$GeV and a $^{32}$S target. Such a measurement would provide a strong proof that dark matter behaves fundamentally different from baryonic matter. We also note that if $\beta$ is actually larger than the 0.1 we have assumed, the required number of events can be reduced significantly.

We have investigated the dependence of the sensitivity with respect to the angular resolution and the threshold energy of the detector, the masses of the WIMP and the target and the background level. The main point is that low energy and high energy nuclear recoils carry more information about the velocity anisotropy, implying a steep calibration curve, while intermediate energy recoils show little sensitivity. Since the calibration curves for low and high energy recoils have positive and negative slope, respectively, an energy-sensitive detector may improve the sensitivity.

\ack The Dark Cosmology Centre is funded by the Danish National Research Foundation.
\section*{References}
\bibliographystyle{unsrt}
\bibliography{directdm}

\end{document}